\documentclass[fleqn,usenatbib]{mnras}

\usepackage{newtxtext,newtxmath}

\usepackage[T1]{fontenc}

\DeclareRobustCommand{\VAN}[3]{#2}
\let\VANthebibliography\thebibliography
\def\thebibliography{\DeclareRobustCommand{\VAN}[3]{##3}\VANthebibliography}

\usepackage{graphicx}	

\def\gsim{\mathrel{\rlap{\lower 4pt \hbox{\hskip 1pt $\sim$}}\raise 1pt
\hbox {$>$}}}
\def\lsim{\mathrel{\rlap{\lower 4pt \hbox{\hskip 1pt $\sim$}}\raise 1pt
\hbox {$<$}}}

\title[An envelope around over-luminous SNe Ia]{Initial Flash and Spectral Formation of Type Ia Supernovae with An Envelope: Applications to Over-luminous SNe Ia}

\author[K. Maeda et al.]
{Keiichi Maeda,$^{1}$\thanks{E-mail: keiichi.maeda@kusastro.kyoto-u.ac.jp}
Ji-an Jiang,$^{2}$
Mamoru Doi,$^{3,4,5}$
Miho Kawabata,$^{1}$
and Toshikazu Shigeyama$^{4}$
\\
$^{1}$Department of Astronomy, Kyoto University, Kitashirakawa-Oiwake-cho, Sakyo-ku, Kyoto, 606-8502. Japan\\
$^{2}$National Astronomical Observatory of Japan, National Institutes of Natural Sciences, 2-21-1 Osawa, Mitaka, Tokyo 181-8588, Japan\\
$^{3}$Institute of Astronomy, Graduate School of Science, The University of Tokyo, 2-21-1 Osawa, Mitaka, Tokyo 181-0015, Japan\\
$^{4}$Research Center for the Early Universe, Graduate School of Science, The University of Tokyo, 7-3-1 Hongo, Bunkyo-ku, Tokyo 113-0033, Japan\\
$^{5}$Kavli Institute for the Physics and Mathematics of the Universe (WPI), The University of Tokyo Institutes for Advanced Study, The University of Tokyo, \\5-1-5 Kashiwanoha, Kashiwa, Chiba 277-8583, Japan
}

\date{Accepted XXX. Received YYY; in original form ZZZ}

\pubyear{2022}

\begin{document}
\label{firstpage}
\pagerange{\pageref{firstpage}--\pageref{lastpage}}
\maketitle

\begin{abstract}
Over-luminous type Ia supernovae (SNe Ia) show peculiar observational features, for which an explosion of a super-massive white dwarf (WD) beyond the classical Chandrasekhar-limiting mass has been suggested, largely based on their high luminosities and slow light-curve evolution. However, their observational features are diverse, with a few extremely peculiar features whose origins have not been clarified; strong and persisting C II lines, late-time accelerated luminosity decline and red spectra, and a sub-day time-scale initial flash clearly identified so far at least for three over-luminous SNe Ia. In the present work, we suggest a scenario that provides a unified solution to these peculiarities, through hydrodynamic and radiation transfer simulations together with analytical considerations; a C+O-rich envelope ($\sim 0.01 - 0.1 M\odot$) attached to an exploding WD. Strong C II lines are created within the shocked envelope. Dust formation is possible in the late phase, providing a sufficient optical depth thereafter. The range of the envelope mass considered here predicts an initial flash with time-scale of $\sim 0.5 - 3$ days. The scenario thus can explain some of the key diverse observational properties by a different amount of the envelope, but additional factors are also required; we argue that the envelope is distributed in a disc-like structure, and also the ejecta properties, e.g., the mass of the WD, plays a key role. Within the context of the hypothesized super-Chandrasekhar-mass WD scenario, 
we speculatively suggest a progenitor WD evolution including a spin-up accretion phase followed by a spin-down mass-ejection phase. 
\end{abstract}

\begin{keywords}
Supernovae: general -- transients: supernovae -- radiative transfer -- circumstellar matter -- white dwarfs
\end{keywords}



\section{Introduction} \label{sec:intro}

There is a general consensus that Type Ia supernovae (SNe Ia) are thermonuclear explosions of a C+O white dwarf (WD), while the progenitor system(s) and explosion mechanism(s) have been actively debated \citep[e.g.,][for a review]{maeda2016}. Either of the `single-degenerate (SD) scenario' \citep[i.e., a WD accreting materials from a non-degenerate companion:][]{whelan1973,nomoto1982a} or the `double-degenerate (DD) scenario' \citep[i.e., a binary WD merger:][]{iben1984,webbink1984} can potentially lead to the thermonuclear runaway, either as a deflagration-triggered explosion initiated deep inside the core of a WD close to the Chandrasekhar-limiting mass ($M_{\rm Ch}$)  \citep[e.g.,][]{khokhlov1991,iwamoto1999,maeda2010a,seitenzahl2013} or a detonation-triggered explosion starting near the surface of a sub-Chandrasekhar-limiting mass ($M_{\rm sub-Ch}$) WD  \citep[e.g.,][]{nomoto1982b,livne1990,woosley1994,shen2009}. 

A peculiar class of over-luminous SNe Ia challenges these scenarios. They were initially identified by high luminosity that requires a large amount of $^{56}$Ni as a power source, broad light curves that can be explained by massive ejecta, and slow expansion velocities seen in spectral lines \citep{howell2006}. These properties are seen in a comprehensive data set for the prototypical over-luminous SN Ia 2009dc \citep{yamanaka2009,silverman2011,taubenberger2011}, which also shows strong C II 6,580 and C II 7,234. The existence of the strong C II lines is now regarded as a striking feature characterizing the over-luminous SNe Ia. Indeed, it has recently been reported that SN 2020esm, as discovered and followed up from an infant phase soon after the explosion, showed very strong C II 6,580 and C II 7,234 and almost no trace of Si II 6,350 in its earliest spectrum \citep{dimitriadis2022}; SN 2020esm then evolved to show a spectrum similar to SN 2009dc at the maximum light. 

The properties summarized above were, at least initially, thought to fit into the `super-Chandrasekhar-mass' ($M_{\rm sup-Ch}$) WD scenario, and therefore they may be referred to as super-Chandrasekhar SN Ia `candidates'. Evolutionary scenarios leading to the formation and explosion of a $M_{\rm sup-Ch}$ WD have thus been considered and proposed by various researchers (see Section 4.4 for further details). However, with an increasing number of similar events, a huge diversity has been noticed within this `over-luminous' class, current classification of which is largely based on the spectral similarity and slow evolution in the light curve; some of them may be just moderately luminous (depending on the treatment of the host extinction which is frequently very uncertain) and/or the velocities seen in the spectral lines are not always slow \citep[e.g.,][]{hicken2007,scalzo2010,chakradhari2014,yamanaka2016,ashall2021,srivastav2022}. 

Indeed, unlike the initial expectation, it has been shown that high-velocity lines (comparable to normal SNe Ia) are more in line with what the $M_{\rm sup-Ch}$ WD scenario predicts \citep{maeda2009_supch,hachinger2012,fink2018}. An interesting, possible solution on this has been proposed; the existence of a massive circumsltellar material (CSM) or envelope attached to the exploding WD. The scenario was initially raised to explain the combination of the slow evolution and high luminosity \citep{noebauer2016} based on the `normal' SN Ia ejecta, as the SN-CSM interaction introduces an additional power and diffusion time-scale. While this scenario has not been rejected in general, at least three counter examples have been recently reported; SN 2020hvf, which also falls into the category of the over-luminous class (while its peak luminosity is just moderately high, with a large uncertainty of the host extinction), showed a sub-day time-scale bright flash within a day of the putative explosion date \citep{jian2021}. Recently added are SNe 2021zny \citep{dimitriadis2023} and 2022ilv \citep{srivastav2022}, which show a similar initial flash; while the sample is still limited, the initial flash is generally (so far always) found for the over-luminous SNe Ia when prompt photometric observations just after the explosion are conducted with sufficiently high sensitivity. If the SN-CSM interaction scenario would be to provide a major power input to the system, such a short time-scale phenomenon must be diluted out, irrespective of the origin of the flash. The initial flash may indeed be common for over-luminous SNe Ia; there are other two over-luminous SNe Ia that show the early-excess in their light curves, LSQ12gpw \citep{firth2015,walker2015,jiang2018} and ASASSN-15pz \citep{chen2019}, while the duration is not well defined for these cases.

The initial, sub-day time-scale flash associated with SN 2020hvf is best explained by a moderately massive ($\sim 0.01 M_\odot$) and compact ($\sim 10^{13}$ cm) `envelope' (see Section 3.1 for further details). This opens up an interesting avenue on possible roles of such an envelope on `spectroscopic' properties; the deceleration due to the interaction may lead to a slow expansion velocity, as well as strong C II lines if the envelope is C+O-rich in its composition \citep{ashall2021}. However, spectral model investigation has been missing. Based on the insight obtained for the nature of the envelope for SN 2020hvf, in this paper we tackle this issue. 

Another striking feature of over-luminous SNe Ia, whose origin has not been clarified yet, is a peculiar late-time behaviour. An extremely rapid decline in the luminosity, which is in conflict with the energy input by the $^{56}$Co decay, was first reported for SN 2006gz, together with the extremely red nature of its late-time spectrum \citep{maeda2009_2006gz}. The increasing sample shows that this behaviour is commonly seen in over-luminous SNe Ia, with a varying degree seen both in the luminosity decline rate and red color \citep{taubenberger2013}. As one possibility, dust formation has been proposed, which could explain both the accelerated luminosity decline and the red spectra in the late phase \citep{maeda2009_2006gz}. 
This has been further demonstrated, in a phenomenological manner, by \citet{taubenberger2013}, based on the correlation between the late-time luminosity decline rate and the degree of the flux suppression in the blue portion of the late-time spectra. In this paper, we investigate the possibility of the dust formation within the cool, shocked-envelope shell. 

The paper is structured as follows. In section 2, methods are described for our simulations, where we adopt a $M_{\rm sup-Ch}$ WD ejecta model attached with different amounts of the envelope; the main focus of the present work is however expected to be insensitive to the inner ejecta properties. In Section 3, we first present photometric properties from our models, then spectroscopic properties after the initial flash. Discussion is given in Section 4 for several implications; a possible geometrical structure of the envelope (Section 4.1), dust formation and expected late-time properties (Section 4.2), limitation of the roles of the envelope to explain the observational properties of over-luminous SNe that will require an additional controlling factor (Section 4.3), a possible evolutionary channel (Section 4.4), and implications for other subclasses including the normal ones (Section 4.5). The paper is closed in Section 5 with a summary of our findings. 


\section{Methods}
The configuration examined in the present work is based on the model for SN 2020hvf presented in \citet{jian2021}, and we restrict ourselves to spherical configuration for the input models (see Section 4.1 for discussion on possible asymmetry). We consider a specific ejecta model with the total mass of $2.1 M_\odot$ (i.e., $M_{\rm sup-Ch}$ WD) and $1.4 \times 10^{51}$ erg for the kinetic energy. The density structure is assumed to be exponential \citep[e.g.,][]{kasen2006,maeda2018}, which mimics typical explosion models \citep[e.g.,][]{nomoto1982a,iwamoto1999}. This ejecta model has three zones in the abundance structure; the $^{56}$Ni-rich zone ($1.6 M_\odot$), the Si-rich zone ($0.3 M_\odot$), and the O-rich zone ($0.2 M_\odot$), from the inner to the outer region. The mass fractions of elements/isotopes in each zone are set to represent characteristic elemental compositions seen in the simulations of explosive burning \citep[e.g.,][]{iwamoto1999}; for example, in the $^{56}$Ni-rich zone, the mass fraction of $^{56}$Ni is assumed to be $0.9$, and thus the ejecta have $1.44 M_\odot$ of $^{56}$Ni. No carbon is included in any of the zones of the ejecta model. 

Attached to this ejecta model is the `envelope' (or confined CSM). This component is assumed to follow the power-law density distribution with the index of $-3$ \citep{piro2016}, which is motivated by the post-merger configuration seen in simulations of binary WD mergers \citep[e.g.,][]{pakmor2012,schwab2012,tanikawa2015} while we do not intend to specify the origin of the envelope. The outer radius is fixed to be $10^{13}$ cm as motivated by the modeling result for SN 2020hvf \citep{jian2021}, and studying the dependence on the envelope radius is postponed to the future \citep[but see][for photometric properties]{piro2016,maeda2018}. The mass of the envelope, $M_{\rm env}$, is a main parameter (including a model without the envelope). The composition in the envelope is divided into carbon and oxygen equally in the mass fractions, with additional solar metal composition added beyond neon up to Fe-peak elements. 

We note that the ejecta properties adopted in the present work is based on the hypothesized $M_{\rm sup-Sch}$ WD scenario, which is however yet to be established. Indeed, the WD mass of $2.1 M_\odot$ corresponds to the uppermost limit of the model sequence of rapidly-rotating WD models computed by \citet{yoon2005}, and the formation of such a massive WD is still highly speculative. It has not been clarified if the $^{56}$Ni mass of $1.44 M_\odot$ as adopted in the present work can be realized in the explosion. We however emphasize that the effect of the envelope as a main interest of the present work is expected to be insensitive to the inner ejecta properties. The present results can thus be applicable irrespective of the nature of the ejecta. 

\begin{figure}
\centering
\includegraphics[width=\columnwidth]{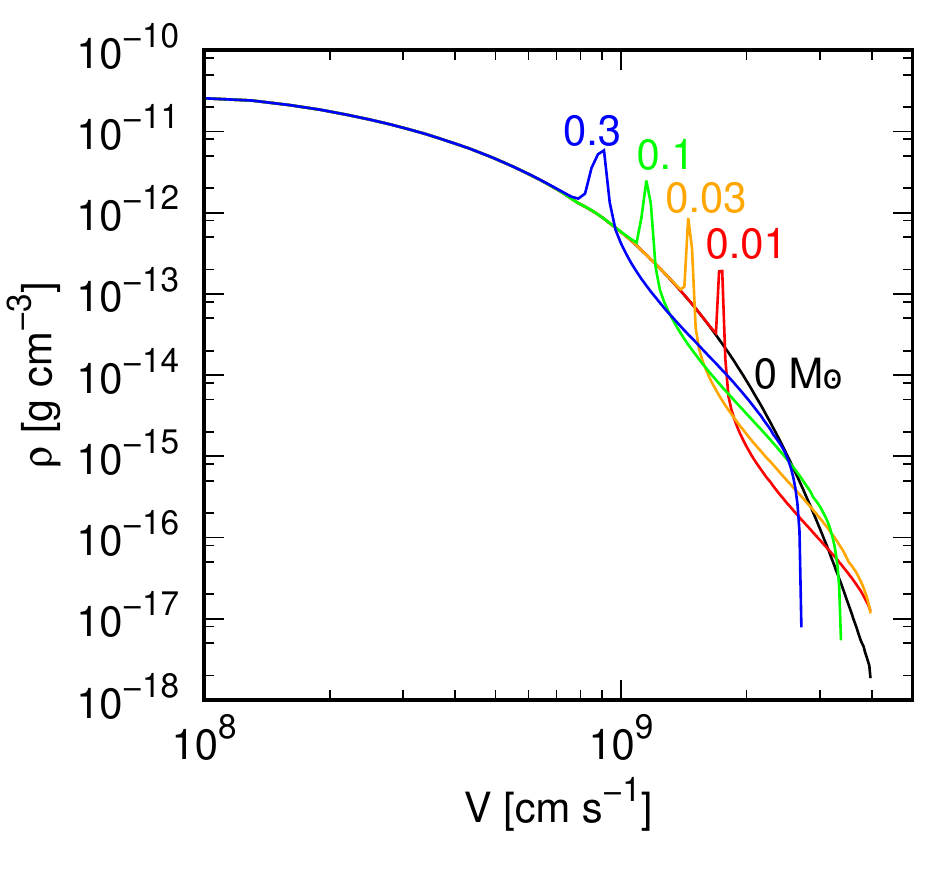}
\includegraphics[width=\columnwidth]{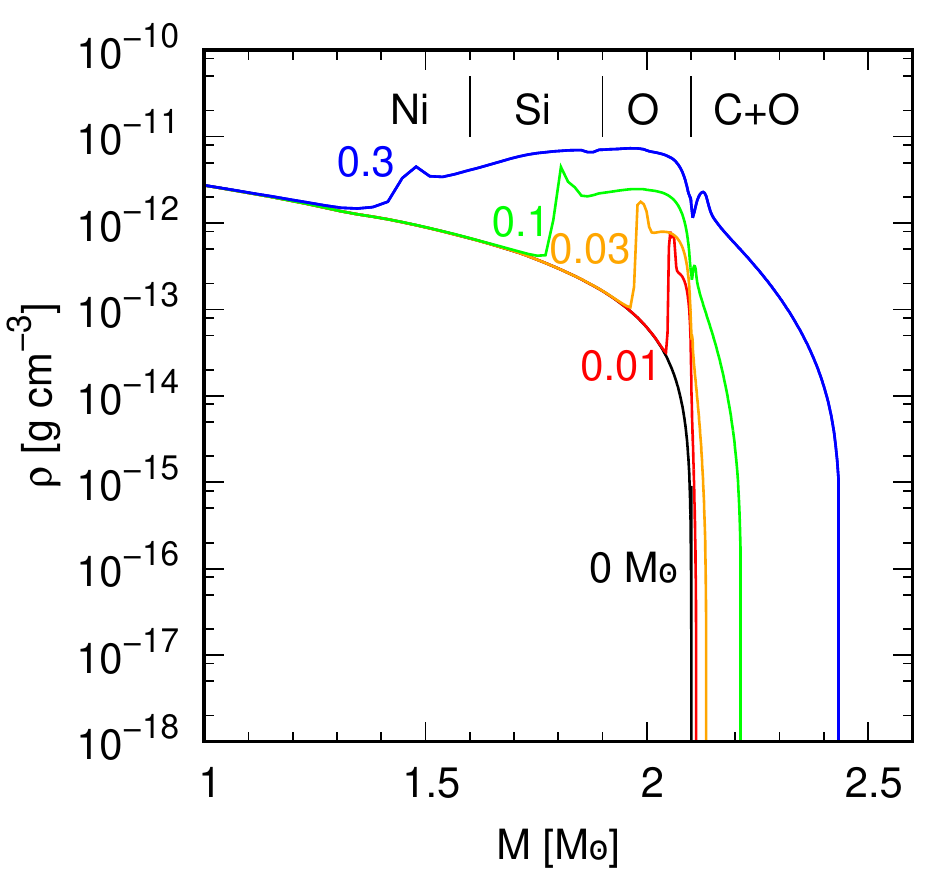}
\caption{The density structure of the ejecta as a result of the interaction, for $M_{\rm env} = 0$ (black), $0.01$ (red), $0.03$ (orange), $0.1$ (green), and $0.3 M_\odot$ (blue), in the velocity coordinate (top) and the mass coordinate (bottom). The distribution of the characteristic burning layers is indicated by the labels in the bottom panel. The density is scaled at 8 days since the explosion. 
}
\label{fig:density}
\end{figure}

As the first step, the hydrodynamical interaction between the ejecta and the envelope is simulated with SNEC \citep[the SuperNova Explosion Code;][]{morozova2015}, in the radiation-hydrodynamic mode. The interaction produces a high-temperature but rapidly cooling `fireball', as is analogous to the cooling-envelope emission in core-collapse SNe. The radiation property in this phase is well described under approximations adopted in SNEC, at least in the $B$ and $V$ bands considered here \citep[e.g.,][]{piro2016}, thus we use the output of the SNEC to compute the multi-band LCs resulting from the interaction and cooling. 

As the second step, we simulate multi-band light curves and spectra in the post-interaction, $^{56}$Ni/Co-heating dominating phase, using HEIMDALL \citep[Handling Emission In Multi-Dimension for spectrAL and Light curve calculations;][]{maeda2006,maeda2014}. It is a Monte-Carlo radiation transfer code, supplemented with $\sim 5 \times 10^5$ bound-bound transitions, as well as free-free and bound-free transitions, using the opacity data taken from \citet{kurucz1995} and the expansion-opacity formalism \citep{karp1977,eastman1993}. After the interaction is over, the density structure is quickly frozen and represented by homologous expansion. The density structure here is mapped from SNEC onto the grids of HEIMDALL. While HEIMDALL can handle multi-dimensional ejecta structure, it is restricted to one-dimensional spherical configuration in the present study. 

\section{Results}

\subsection{Photometric Properties}
\begin{figure}
\centering
\includegraphics[width=\columnwidth]{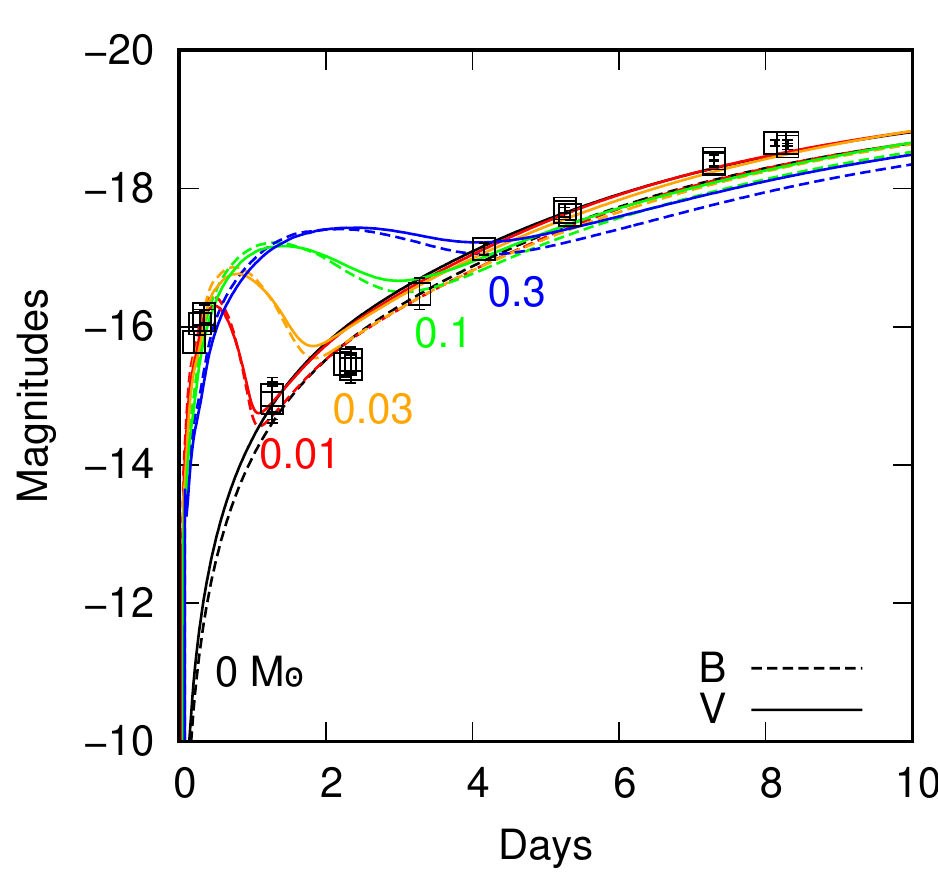}
\caption{Early light curves for the models with different amounts of the envelope; $M_{\rm env} = 0$ (black), $0.01$ (red), $0.03$ (orange), $0.1$ (green), and $0.3 M_\odot$ (blue), in the $B$ band (dashed) and $V$ band (solid); a model with a larger amount of the envelope reaches to a brighter peak at a later epoch in the `initial flash'. Also shown for comparison is the light curve of SN 2020hvf \citep[black squares;][]{jian2021}; the data were taken without a filter, the bandpass of which is comparable to a combination of the Pan-STARRS $g$ and $r$ filters.}
\label{fig:lc_early}
\end{figure}

Fig. \ref{fig:density} shows the `frozen' density structure after the interaction. Note that the hydrodynamical interaction itself is completed in time-scale of $\sim 0.1$ days. There is further modification of the outermost density structure due to the leakage of radiation, i.e., the cooling process, but this effect quickly becomes ineffective in the first few days. 

It is seen that most of the envelope material is swept up to a narrow velocity range, forming a dense shell. The shell consists of the shocked ejecta material and the envelope, with the comparable masses. The shell velocity is lower for a larger amount of the envelope, naturally expected as the effect of deceleration though momentum conservation; the velocity is $\sim 17,000$ km s$^{-1}$ for $M_{\rm env} = 0.01 M_\odot$ and is decreased to $\sim 9,000$ km s$^{-1}$ for $M_{\rm env} = 0.3 M_\odot$.

Fig. \ref{fig:lc_early} shows the earliest-phase light curves including the `shock-cooling' emission. The basic behaviours of the initial flash follow predictions from analytical consideration \citep{maeda2018}: (1) The time-scale is longer (i.e., the peak date is later) for a model with larger $M_{\rm env}$ due to larger diffusion time-scale within the shell. (2) The peak luminosity is higher for larger $M_{\rm env}$ due to a larger amount of the energy dissipation (due to the deeper penetration of the reverse shock). It is however limited within $\sim 1.5$ mag difference even for a large variation of $M_{\rm env}$ between $0.01$ and $0.3 M_\odot$ since it is compensated by larger diffusion time for larger $M_{\rm env}$. Indeed, the luminosity is expected to be more sensitive to the outer radius of the envelope that controls the amount of the adiabatic expansion/cooling. Thanks to the strong dependence of the peak date on $M_{\rm env}$ and that of the peak luminosity on $R_{\rm env}$, the properties of the envelope can be tightly constrained if observational data are available; $R_{\rm env}$ in the present work is set to $10^{13}$ cm by this argument taking SN 2020hvf as a reference case, for which the initial flash was detected \citep{jian2021} and thus the model can be calibrated. Indeed, it is seen in Fig. \ref{fig:lc_early} that the model with $R_{\rm env} = 10^{13}$ cm and $M_{\rm env}=0.01 M_\odot$ reproduces the `initial flash' of SN 2020hvf as reported by \citet{jian2021}. 

\begin{figure}
\centering
\includegraphics[width=\columnwidth]{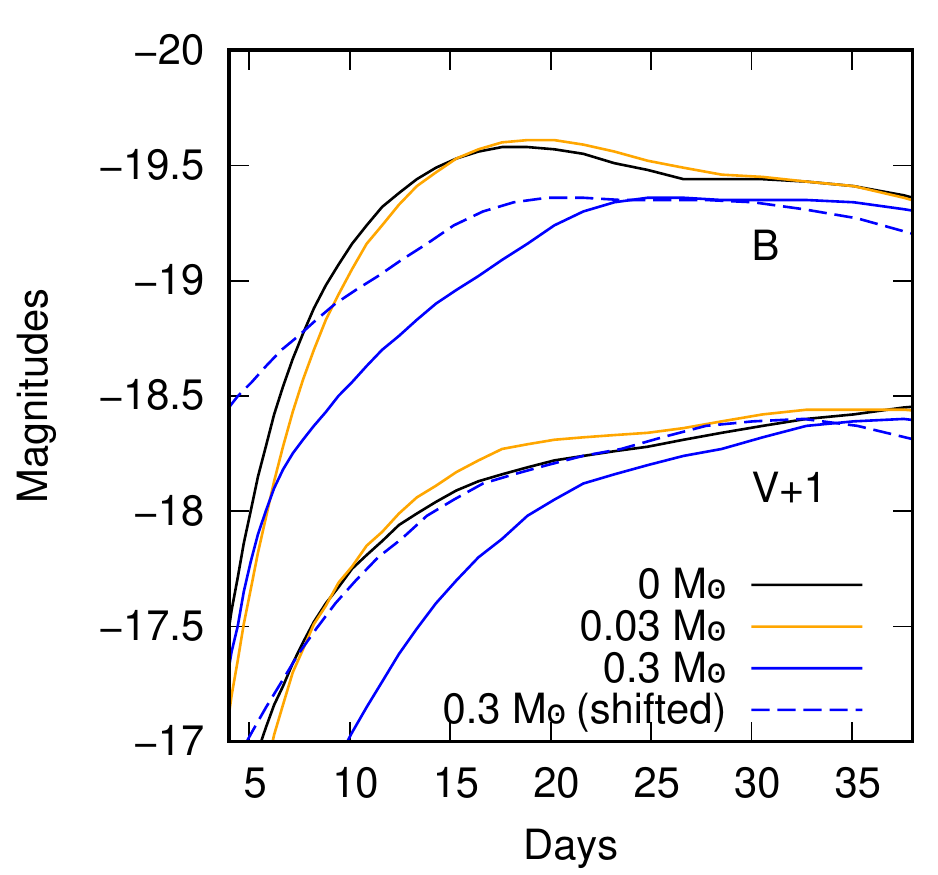}
\caption{$^{56}$Ni-powered light curves for the models with different amounts of the envelope; $M_{\rm env} = 0$ (black), $0.03$ (orange), and $0.3 M_\odot$ (blue), in the $B$ band (upper) and $V$ band (+1 mag; lower). For a demonstration purpose, we also show the model with the envelope mass of $0.3 M_\odot$ but with the days since the explosion artificially shifted by $-5$ days (dashed).
}
\label{fig:lc_max}
\end{figure}

After the initial flash (i.e., cooling emission), the light curve is powered by the decay of $^{56}$Ni to $^{56}$Co then to $^{56}$Fe. Fig. \ref{fig:lc_max} shows the $B-$ and $V-$ band light curves (as computed with detailed opacities in the second step with HEIMDALL). The difference is (only) marginally discerned between models with $M_{\rm env} = 0$ and $0.03 M_\odot$, which is difficult to distinguish from the photometric evolution alone in the post flash phase. The massive envelope (e.g., $M_{\rm env} = 0.3 M_\odot$) creates a noticeable trace in the light curve evolution; the rising is substantially delayed due to the additional diffusion in the swept-up shell \cite[see also][]{noebauer2016}. We however note that the post-flash light curve evolution alone may not be a smoking gun for the existence of the massive envelope. For example, without a priori knowledge of the explosion date, the $V$-band light curve may be nearly indistinguishable from a model without the envelope (Fig. \ref{fig:lc_max}). In addition, other effects, e.g., mixing extent of $^{56}$Ni toward the surface, may also affect the light curve evolution \citep[e.g.,][]{magee2020}. 

\subsection{Spectroscopic Properties}

\begin{figure*}
\centering
\includegraphics[width=\columnwidth]{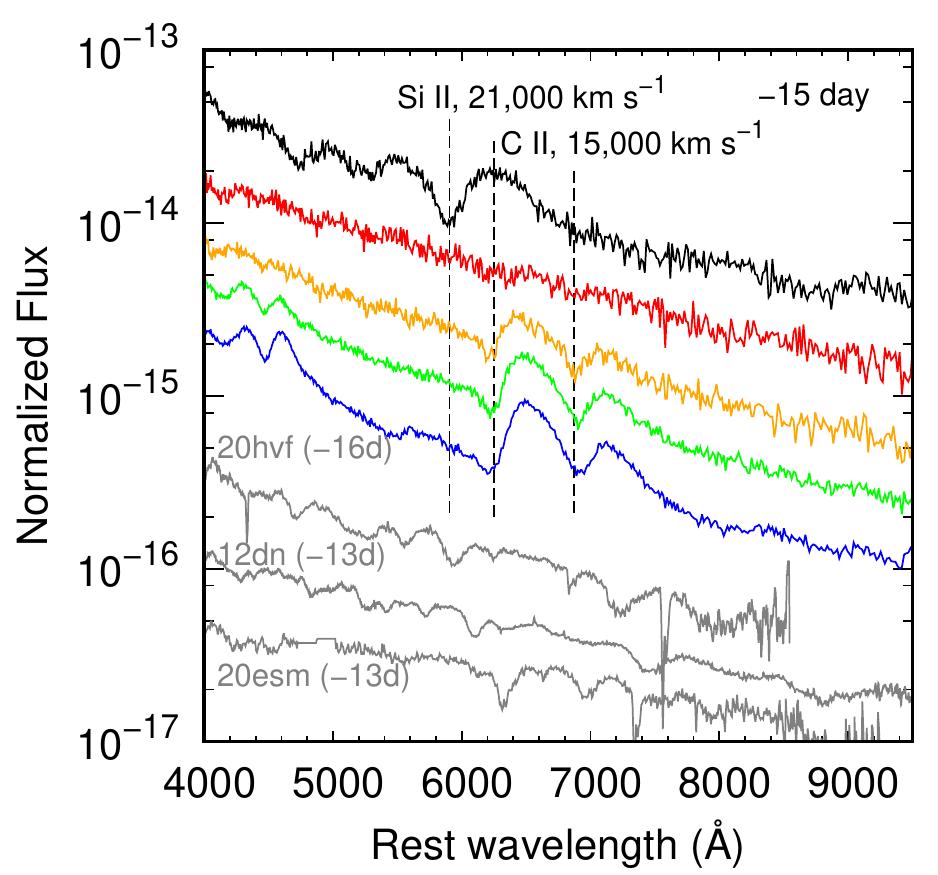}
\includegraphics[width=\columnwidth]{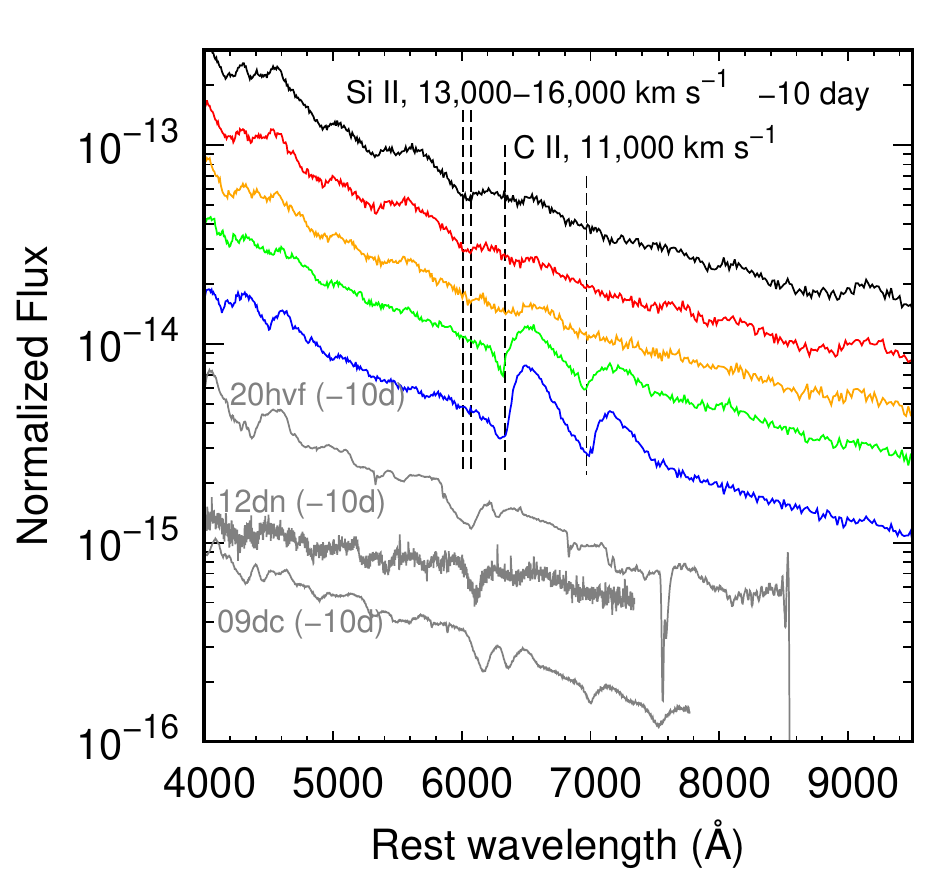}\\
\includegraphics[width=\columnwidth]{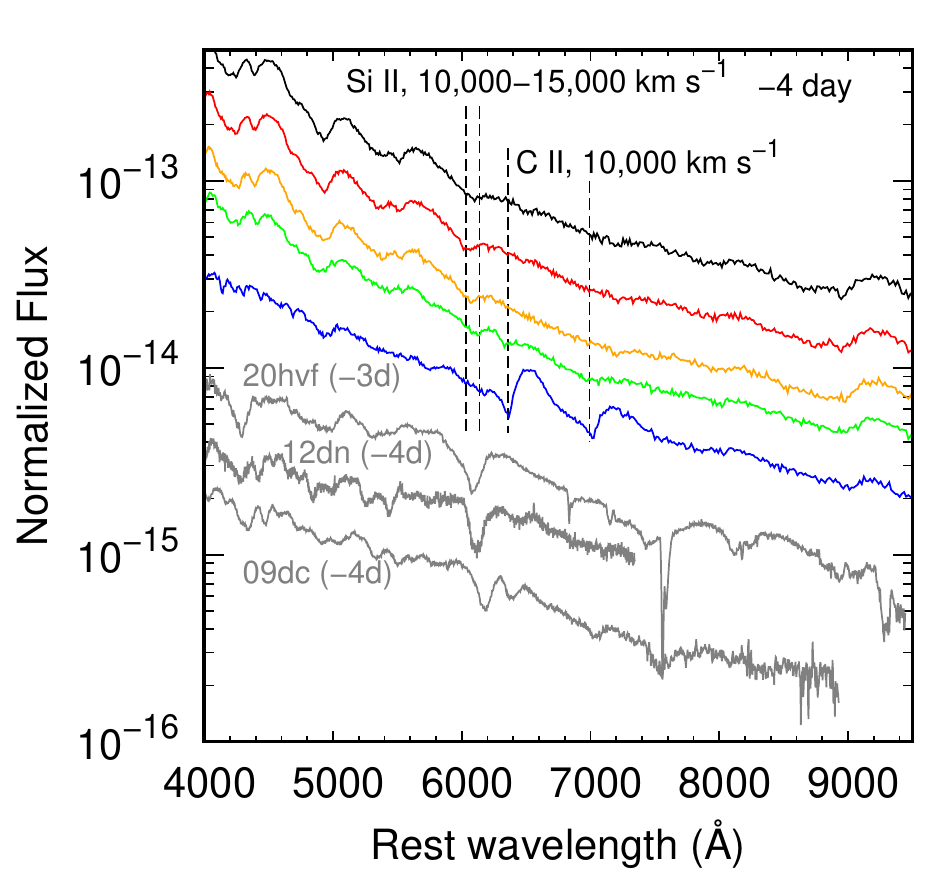}
\includegraphics[width=\columnwidth]{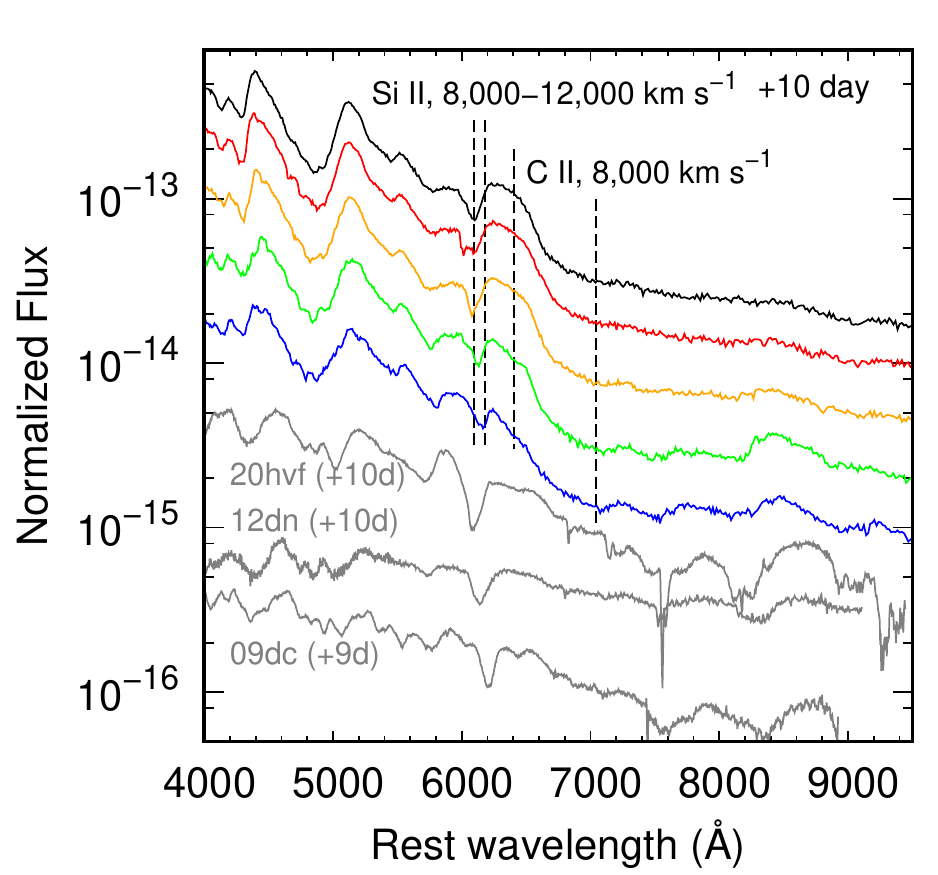}
\caption{The synthetic spectra on days $\sim -15$, $-10$, $-4$, and $+10$ days since the $B$-band maximum. The model spectra are shown for the one without the envelope (black), with the envelope of $0.01$ (red), $0.03$ (orange), $0.1$ (green), and $0.3 M_\odot$ (blue) (from top to bottom). For comparison, spectra of the following SNe are shown; SNe 2020hvf (-16 days), 2012dn (-13 days) and 2020esm (-13 days) for comparison at $\sim -15$ days; SNe 2020hvf (-10 days), 2012dn (-10 days) and 2009dc (-10 days) at $\sim -10$ days; SNe 2020hvf (-3 days), 2012dn (-4 days) and 2009dc (-4 days) at $\sim -4$ days; SNe 2020hvf (+10 days), 2012dn (+10 days) and 2009dc (+9 days) at $\sim +10$ days. The sources of the observational data are the following; \citet{taubenberger2011,parrent2016,stahl2020,tucker2020,jian2021}.
}
\label{fig:spec}
\end{figure*}

Fig. \ref{fig:spec} shows synthetic spectra for the models with different amounts of $M_{\rm env}$, at different epochs. Note that we focus on the post-flash phases as our spectral-synthesis simulations do not apply to the initial flash phase. We also show observed spectra of some `over-luminous' SNe Ia at similar epochs\footnote{The observational data are taken through the WISeRep \citep{yaron2012}; 
\url{https://www.wiserep.org/} .}. We however note that they are shown for qualitative comparison, and detailed and quantitative comparison is beyond the scope of the present work. 

Without the envelope, the spectrum at $\sim -15$ days is characterized by strong Si II $\lambda$6,355, to be classified as an SN Ia. With only $0.01 M_\odot$ of the C+O-rich envelope attached, the infant spectrum becomes strikingly different; the model shows a featureless spectrum. This property stems from the photosphere being formed within the swept-up envelope; assuming a singly-ionized or doubly-ionized condition within the envelope, we estimate that the electron-scattering optical depth within the envelope on day 3 since the explosion to be in the range from $1.5$ to $3 (M_{\rm env} / 0.01 M_\odot$) (see Section 4.2). For $M_{\rm env} \gsim 0.03 M_\odot$, the spectrum shows clear C II lines ($6,580$\AA\ and $7,234$\AA), with increasing strength for larger $M_{\rm env}$. The velocities at the absorption minima of the C II lines also decrease for larger $M_{\rm env}$. We however note that the velocity decrease here does not quantitatively trace the decrease in the `shell' velocity for an increasing envelope mass; the shell velocity decreases from $\sim 17,000$ k ms$^{-1}$ for $M_{\rm env} = 0.01 M_\odot$ to $\sim 9,000$ km s$^{-1}$ for $M_{\rm env} = 0.3 M_\odot$ (Fig. \ref{fig:density}), while the difference in the C II absorption minima in the synthetic spectra is much smaller. The spectral formation in these envelope models at $\sim -15$ days indeed takes place above the `shell', in the region with steep density gradient toward the surface of the shocked envelope. Still, the overall decreasing tendency of the absorption-line velocities are seen for a more massive envelope, which is due to the overall lower velocity of the (shocked) C+O-rich envelope for larger $M_{\rm env}$. 

The spectra dominated by the C II lines without the Si II lines, as synthesized with $M_{\rm env} \gsim 0.03 M_\odot$, do not represent `classical' SN Ia spectra. This feature indeed provides a striking similarity to the spectrum of the over-luminous SN Ia 2020esm \citep{dimitriadis2022} taken in such an infant phase (Fig. \ref{fig:spec}). Note that the velocities seen in the models are higher than observed, but we emphasize that our reference model has been constructed with SN 2020hvf as a reference, which shows the highest-velocity absorptions among the over-luminous class (see Fig. \ref{fig:spec}). Also seen within the present model framework is that the formation of the Si II and C II lines are mutually exclusive, without having model spectra showing both features in the same snapshot. At this earliest phase, even with the envelope mass as small as $0.01 M_\odot$, the photosphere forms within the C+O-rich layer removing a trace of the Si II lines. This issue will be further addressed in Section 4.1. 

The properties of the spectra at advanced epochs can be interpreted along the same line. On day -10, the models with $M_{\rm env} = 0.01$ and $0.03 M_\odot$ show the spectra nearly identical with the one that does not have the envelope. The photosphere has now receded into the Si-rich ejecta, leading to the formation of the Si II lines. The indication of the Si-rich ejecta layer is seen in the Si II absorption minima in these models (i.e., $\sim 16,000$ km s$^{-1}$ for $M_{\rm env} = 0$ while $\sim 13,000$ km s$^{-1}$ for $M_{\rm env} = 0.03 M_\odot$). We note that a weak trace of the C II is indeed seen in the model with $M_{\rm env} = 0.03 M_\odot$, which exhibits a composite spectrum of the Si II and C II lines in the small time window under the present model framework. The models with a massive envelope ($M_{\rm env} = 0.1$ and $0.3 M_\odot$) keep showing the C-rich spectra with the C II line velocity substantially decreased due to the recession of the photosphere and the line-forming region; the velocity here is $\sim 11,000$ km s$^{-1}$, which is close to the shell velocity for $M_{\rm env} = 0.1 M_\odot$ but is still substantially higher for $M_{\rm env} = 0.3 M_\odot$. The sample of over-luminous SNe Ia around this phase show strong C II lines, whose strengths are comparable to that of Si II 6,350 for an exceptional case of SN 2009dc but is weaker than the Si II for the other SNe Ia. 

The spectra further evolve, and the trace of the C II is almost completely gone for $M_{\rm env} \lsim 0.03 M_\odot$ around the maximum light. Now the model with $M_{\rm env} = 0.1 M_\odot$ shows the composite-type spectrum, following further recession of the photosphere inside the Si-rich layer of the ejecta. The Si II velocities, if visible (except for $M_{\rm env} = 0.3 M_\odot$), show a decreasing trend for larger $M_{\rm env}$; this behaviour originates in deceleration of the Si-rich ejecta due to the interaction with the envelope. It is seen that the Si II-forming region at this phase is just behind the shell (see Figs. \ref{fig:density} and \ref{fig:spec}). The decrease in the C II strength from the earlier epochs is seen for a fixed model parameter, which qualitatively explains a general trend seen in the evolution of the sample of over-luminous SNe Ia. 

In the post-maximum phase on day +10, even the model with $M_{\rm env} = 0.3 M_\odot$ starts showing the Si II; the photosphere is now below the shell in all the models. Again, lower line velocities for larger $M_{\rm env}$ are discerned. The models with $M_{\rm env} \gsim 0.1 M_\odot$ still shows a trace of the C II lines together with the Si II. The sample of over-luminous SNe Ia keep showing the decrease in the C II strength; for SNe 2020hvf and 2012dn, the C II is no more clearly identified, while it is still clearly detected for SN 2009dc. These features are qualitatively consistent with the model expectations (while noting that the model sequence here is based on the investigation of the spectral evolution of SN 2020hvf, and is not tuned to other over-luminous SNe Ia); SN 2009dc corresponds to a model with a massive envelope, and SNe 2020hvf and 2012dn correspond to a model with a less massive envelope.

\section{Discussion}

Out findings can be summarized as follows: 
\begin{enumerate}
\item The energy stored in the interaction between the ejecta and the envelope creates an initial flash due to the shock-cooling process, with the increasing duration for a more massive envelope. 
\item The duration of the initial flash ranges between $\sim 0.5$ and $\sim 4$ days for $M_{\rm env} \sim 0.01$ to $0.3 M_\odot$. 
\item The swept-up C+O-rich envelope can produce C II 6,580 and C II 7,234. 
\item For $M_{\rm env} \gsim 0.03 M_\odot$, they initially show spectra dominated by the C II lines with little trace of the Si II lines. For $M_{\rm env} \sim 0.01 M_\odot$, the initial spectrum looks like featureless without clear Si II nor C II lines. 
\item The strength of the C II becomes weaker as time goes by, and eventually the spectra turn into the `SN Ia' spectra as characterized by strong Si II 6,350. 
\item The strength of the C II lines is weaker for a smaller amount of the C+O-rich envelope at a given epoch. Accordingly, the transition of the characterizing lines from the C II to the Si II takes place earlier for a less massive envelope. 
\item The composite spectra, showing both of the Si II and C II lines, are found in a relatively brief time window. For a smaller amount of the envelope, this `composite' phase is shorter. 
\item The velocities at the absorption minima of the Si II are lower for a more massive envelope, reflecting the decelerated Si-rich ejecta. The similar tendency is also seen in the C II velocities. 
\end{enumerate}

The features summarized above can explain some of the characteristic features and evolution seen in over-luminous SNe Ia, especially the strong C II lines early on (including the spectra dominated totally by these lines as seen in SN 2020esm) and their subsequent disappearance. At the same time, some properties are yet to be explained. Our model sequence is based on the model for SN 2020hvf in its construction, and its application to other over-luminous SNe Ia is limited. Still, even with this caveat bared in mind, there is one critical issue; our present model does not explain the spectra with long-lasting strong Si II {\em and} C II lines, especially in the very early phase. This problem would not be solved by simply changing the eject and envelope models within the present model framework. This is the issue we will address in Section 4.1. 

Another striking feature of over-luminous SNe Ia is their late-time behaviour, showing an accelerated luminosity decline and a red color (or suppression of flux in the shorter wavelengths), as introduced in Section 1. Under the dust formation scenario as one possibility \citep{maeda2009_2006gz,taubenberger2013}, the present model provides an interesting possibility; the dust may be formed within the dense and cool C+O-rich shell. This topic will be further investigated in Section 4.2. 

Another unresolved issue is that the present model would not explain all the major diversities seen among over-luminous SNe Ia by simply changing the properties of the envelope (as repeatedly mentioned as a caveat). It is then important to address what could be attributed to the diversity in the envelope properties, and what should require additional sources of the diversity. This is the issue to be addressed in Section 4.3. 

One possible scenario frequently discussed for the over-luminous SNe Ia is an explosion of a `super-Chandrasekhar-mass' WD. Putting the present findings and implications all together, we will discuss a possible evolutionary channel leading to such a configuration in Section 4.4. 

In the present work, we focus on the possibility that the existence of the envelope may explain some of the major properties seen in over-luminous SNe Ia. This is motivated by the observational properties of SN 2020hvf, which shows the expected features of such an envelope both in photometric and spectroscopic properties (Section 3). However, the applicability of the present analysis is not limited to this special class. While the detailed comparison requires the ejecta model(s) tuned to each subclass, general and qualitative discussion on the existence of a similar envelope is possible. This is a topic in Section 4.5.

\subsection{Asymmetry in the envelope structure?}

The persistent coexistence of the Si II and C II lines is difficult to explain within the present model framework. The models predict the C II-dominating spectrum and the Si II-dominating one showing up in two distinct characteristic phases, and the intermediate `transition' phase showing the composite spectrum lasts only for a limited time window. This stems from the evolution of the photosphere, starting from the formation within the C+O-rich envelope then followed by the recession to reach to the Si-rich ejecta. 

\begin{figure}
\centering
\includegraphics[width=1.15\columnwidth]{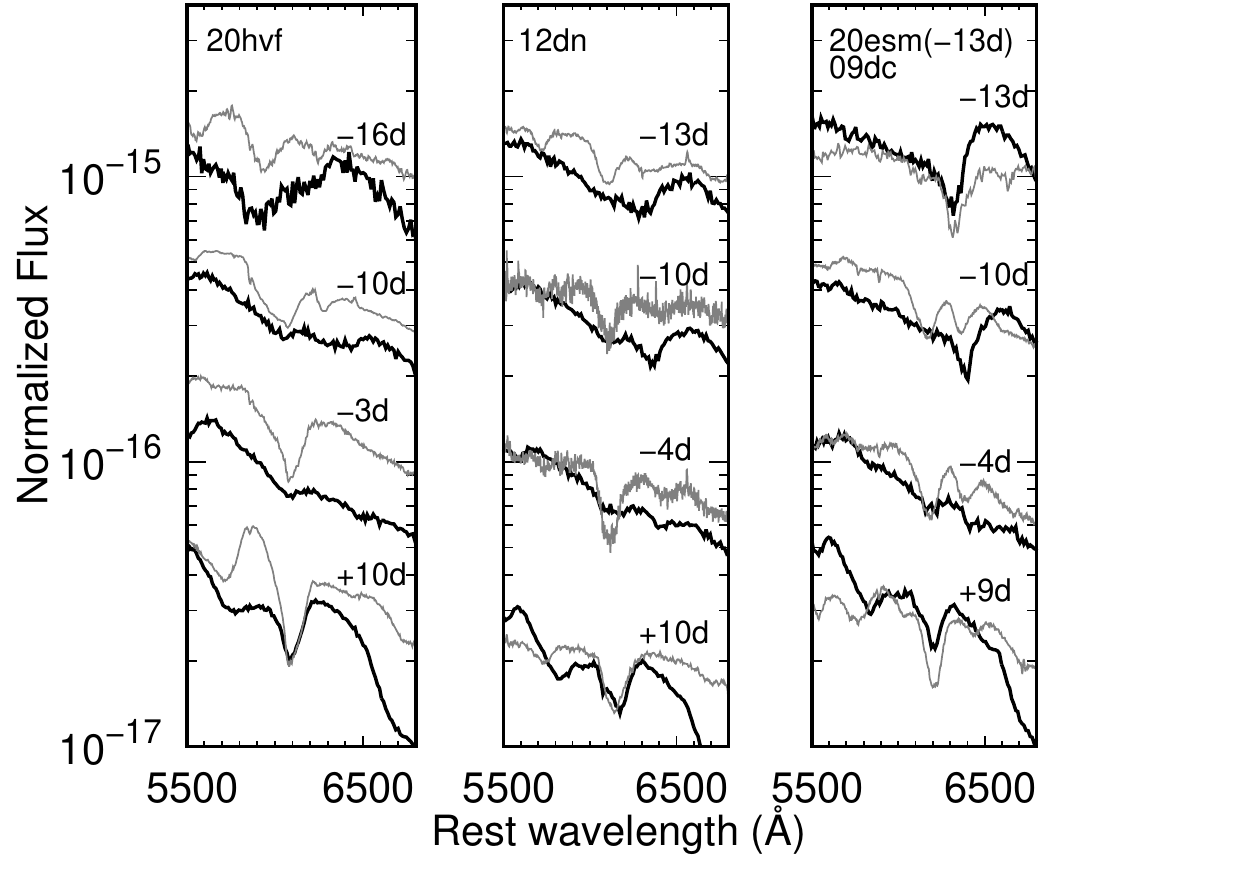}
\caption{Comparison between the spectral evolution (thin-gray) of SNe 2020hvf (left), 2012dn (middle), and 2020esm/2009dc (right), and the hybrid models with different amounts of the C+O envelope (thick-black). The models are the following; \textit{(left)} the contribution by the model without the envelope and that with $M_{\rm env} = 0.03 M_\odot$ are equally added; \textit{(middle)} the contribution by the model with $M_{\rm env} = 0.01 M_\odot$ and that with $M_{\rm env} = 0.1 M_\odot$ are equally added, with an additional redshift of $3,000$ km s$^{-1}$ applied to the model; \textit{(right)} the model with $M_{\rm env} = 0.1 M_\odot$ only, with an additional redshift of $1,500$ for the comparison to SN 2020esm and $3,900$ km s$^{-1}$ for the comparison to SN 2009dc (Tab. \ref{tab:velocity}). 
}
\label{fig:spec_expand}
\end{figure}

One possible solution to remedy this issue is a possible asymmetry in the envelope distribution, so that the Si-rich ejecta are freely expanding toward a specific direction while encountered by the C+O-rich envelope in the other direction. This may happen either in a disc-like or (large-scale) clumpy structure in the envelope distribution. The former could be indeed a natural configuration within the context of the `super-Chandrasekhar-mass' WD progenitor model involving the support by the centrifugal force \citep{uenishi2003,yoon2005}; the nearly break-up velocity required in such a scenario may create a centrifugally-supported envelope in a disc shape \citep{paczynski1991} (see Section 4.4 for further discussion). 

As a rough description of such an asymmetric system, we may introduce a combination of two models with different amounts of the envelope while keeping the same ejecta properties. Fig. \ref{fig:spec_expand} shows this investigation, for demonstration purpose. Compared with the spectral evolution of SN 2020hvf is a combination of the models with $M_{\rm env} = 0$ and $M_{\rm env} = 0.03 M_{\odot}$, assuming the equal contributions from the two at all the epochs. Since the model without the envelope was constructed in a way to roughly reproduce characteristic features of SN 2020hvf (but noting that the model was not tuned to find a detailed fitting to the data), it traces the evolution of the Si II velocity reasonably well. The C+O-rich envelope was not included in the spectral calculations presented in \citet{jian2021}, while it was considered in the photometric properties; they estimated $M_{\rm env} \sim 0.01 M_\odot$ to explain the luminosity and duration of the initial flash (see also Fig. \ref{fig:lc_early}). The combined model here roughly explains the evolution of C II 6,580, including its existence on day -16 and subsequent weakening in its strength. If we would take an average of the envelopes in the two components with the relative weight used in the comparison, it is $0.015 M_\odot$; it is expected that the light curve evolution including the initial flash can also be explained, especially if one would further fine tune the radius of the envelope. 

For the other over-luminous SNe Ia, such exercise should be taken as demonstration, since the ejecta model is not tuned to represent different SNe Ia (see also Section 4.3). To compare with the properties of SN 2012dn, we consider a combination of the models with $M_{\rm env} = 0.01 M_\odot$ and $0.1 M_\odot$ in Fig. \ref{fig:spec_expand}, again with equal contributions. For the reason mentioned above, we artificially introduce an additional redshift of 3,000 km s$^{-1}$ to the model spectra so that the Si II line velocity evolution is roughly explained (Tab. \ref{tab:velocity}; see Section 4.3 for further discussion). Qualitatively, a rough description of the observed spectral evolution is reached by this model; the persistent (and weakening) C II line up until the maximum light. 

We then consider a combined evolution of SNe 2020esm and 2009dc, given the similarity of the maximum-light spectra for the two SNe. The model shown here is indeed a single model with $M_{\rm env} = 0.1 M_\odot$. The additional redshift is introduced for the model spectra, 1,500 km s$^{-1}$ on day -13 (when compared to SN 2020esm) and 3,900 km s$^{-1}$ on the other epochs (when compared to SN 2009dc), The model here fails to explain the emergence of the Si II on day -10, but otherwise basic properties are roughly and qualitatively explained. Especially noticeable is the strong C II lines and the absence of the Si II lines on day -13, which is a characteristic feature of SN 2020esm in its infancy. 

Note that introducing the bulk redshift produces `unrealistic' shift in the emission peak of the P-Cygni profile, and only the absorption minimum should be compared between the `shifted' model spectra and the observed spectra; the difference in the expansion velocity should affect the absorption component of the P-Cygni profile leaving the emission peak unchanged. This artifact is for example seen in the comparison with SN 2009dc in Fig. \ref{fig:spec_expand}, where the emission peaks in the shifted model are too red as compared to the observed peak positions.

In summary, while we do not intend to provide detailed and quantitative fiting to the observed spectra of a sample of over-luminous SNe Ia, the characteristic properties of their spectral evolution can be explained, at least at a qualitative level. The relative contributions from the models with and without an envelope are taken to be about fifty--fifty (with a large uncertainty) for SN 2020hvf. This might translate to the `covering fraction' of the C+O-rich envelope being $\sim 50$\%, while this should depend on the detailed configuration where the viewing angle can also be an important factor (see Also Section 4.2). A similar argument applies to SN 2012dn, for which the model with $M_{\rm env} = 0.01 M_\odot$ and that with $M_{\rm env} = 0.1 M_\odot$ are equally added. 

The result here provides an interesting avenue for furure observational strategy -- polarization observations especially in the infant phase. So far only two over-luminous SNe Ia, 2007if and 2009dc, have reported polarization, both showing a low polarization level around the maximum light \citep{tanaka2010,cikota2019}. The model presented here for SN 2009dc, i.e., one component model, indeed does not require strong asymmetry, and therefore there is no tension to the polarization constraint. Further, we note that the photosphere has already receded relatively deep into the ejecta, and thus a trace of the asymmetry in the envelope may already be weakened in the maximum-light phase. The consideration here suggests an importance of the very early-phase polarization observation; we predict a high level of the polarization for over-luminous SNe Ia, especially for those showing the composite-type spectra (showing both the Si II and C II) already in the infant phase. 

\subsection{Implications for late-time evolution; dust formation in the shell?}

The over-luminous SNe Ia tend to show peculiar properties in their late-time evolution ($\gsim 100$ days). Some of them show extremely rapid decline in the late-time light curve evolution. These SNe show very red nebular spectra, where [Fe III] and [Fe II] clusters in a blue portion of the spectra as the strongest features in normal SNe Ia are not clearly detected \citep{maeda2009_2006gz}. One of possible origins suggested so far is dust formation within a high-density C+O-rich region \citep{maeda2009_2006gz}. Indeed, the level of the accelerated decline in the late-time light curves and the degree of the late-time spectral blue-side suppression are generally correlated, and they could be consistently explained by the same `reddening' mechanism \citep{taubenberger2013}. In this section, we investigate a possibility as to whether the dust can be formed within the dense and cool shell, i.e., shocked C+O-rich envelope, within the present model framework. 

We may assume that the shocked envelope is confined within a shell with its thickness being $\sim 10$\% of its radius (e.g., the shell width of $\sim 1,000$ km s$^{-1}$ as compared to the shell velocity of $\sim 10,000$ km s$^{-1}$ in the model with $M_{\rm  env} = 0.3 M_\odot$ as shown in Fig. \ref{fig:density}). Then, the average density is estimated as follows; 
\begin{equation}
\rho_{\rm env} \sim 2.5 \times 10^{-16} 
  \left(\frac{M_{\rm env}}{0.1 M_\odot}\right)
  \left(\frac{V_{\rm shell}}{10,000 \ {\rm km s}^{-1}}\right)^{-3} 
  \left(\frac{t}{100 \ {\rm days}}\right)^{-3} \ {\rm g} \ {\rm cm}^{-3} \ .
\end{equation}
It is seen in Fig. \ref{fig:density} that this provides a reasonable estimate. To estimate the temperature evolution of the shell, we may equate the $\gamma$-ray heating rate and the cooling rate by [O I]6,300 \& 6,363. The situation here is analogous to a C/O zone in core-collapse SNe, for which [O I]6,300 \& 6,363 becomes a dominant coolant when the temperature drops below $\sim 5,000$ K \citep{jerkstrand2015}. By omitting other coolants, this indeed provides a conservative estimate for the temperature decrease; with other coolants included, the dust formation will further be accelerated. The $^{56}$Co-decay $\gamma$-ray luminosity is described as 
\begin{equation}
  L_{\gamma} \sim 1.5 \times 10^{43} 
    \left(\frac{M_{\rm 56Ni}}{M_\odot}\right) 
    \exp\left(-\frac{t}{113.5 \ {\rm adys}}\right) \ {\rm erg} \ {\rm s}^{-1} \ .
\end{equation}
The optical depth to the $^{56}$Co decay gamma-rays can be approximated as follows  \citep[e.g.,][]{maeda2003}; 
\begin{equation}
  \tau_{\gamma} \sim 0.005 \left(\frac{M_{\rm env}}{0.1 M_\odot}\right)
  \left(\frac{V_{\rm shell}}{10,000 \ {\rm km s}^{-1}}\right)^{-2} 
  \left(\frac{t}{100 \ {\rm days}}\right)^{-2} \ .
\end{equation}

The cooling process may be represented by [O I]6,300 \& 6,363 to conservatively estimate the possibility of the dust formation (see above). For a representative case with $M_{\rm env} = 0.1 M_\odot$, the oxygen fraction of 50\%, $V_{\rm shell} = 10,000$ km s$^{-1}$, and $t = 100$ days, the number density of oxygen is $\sim 5 \times 10^{6}$ cm$^{-3}$. This is close to the critical density for the [O I] transitions, and we may assume the Local-Thermodynamic Equilibrium (LTE) condition, for the first-order estimate; this may introduce an error of a factor of about two in the cooling rate, which then translates into the error of $\sim 20$\% in the equilibrium temperature as is sufficient for our order-of-magnitude estimate. The cooling rate (for the entire shell) is then approximated by the following \citep[e.g.,][]{uomoto1986,maeda2022}; 
\begin{equation}
  \Lambda \sim 7.5 \times 10^{40} f_{\rm [O I]}^{-1}\exp\left(\frac{-22,860 \ {\rm K}}{T}\right)
  \left(\frac{M_{\rm env}}{0.1 M_\odot}\right) \ {\rm erg} \ {\rm s}^{-1},
\end{equation}
where $f_{\rm [O I]}$ is the fraction of energy channeled through the [O I], and $T$ is the electron temperature within the shell. Here, the mass fraction of oxygen is assumed to be 50\%. 

In the late phase of interest here, $\tau_{\gamma} \ll 1$, so that we can approximate the heating-cooling balance by $L_{\gamma} \tau_{\gamma} \sim \Lambda$. Note that the envelope mass then vanishes, and thus the following consideration is essentially independent from the envelope mass \citep[but noting that a sufficiently high density will be required to form molecules and then dust grains:][]{nozawa2011}. A critical temperature to consider first is $T \sim 5,000$ K, at which it is likely that the CO molecules start forming abundantly within the C+O layer \citep{hoeflich1995}. For our fiducial values (i.e., the scales for the parameters in the above equations, with $f_{\rm [O I]} = 0.3 - 1$), the electron temperature drops to $\lsim 5,000$K in 200 - 300 days after the explosion. Once the CO molecules form, we expect that the cooling is accelerated, and rapidly drops to $\sim 2,000$K at which the C dust grains start forming; the density in the shell, even for small $M_{\rm env}$, will be sufficiently high to have the rapid dust formation at the low temperature, given that the standard SN Ia ejecta model without the CSM interaction already has a sufficiently high density for carbon dust formation \citep[e.g.,][]{nozawa2011}. 

The dust grains, if formed, would affect or even control the late-time observational behaviours \citep{maeda2009_2006gz,taubenberger2013}. Denoting the mass fraction of the carbon grains as $X_{\rm d}$, which may be an order of 0.1 (e.g., $X_{\rm d} = 0.125$ if the mass fractions of C and O are 0.5:0.5, all the oxygen is locked into CO molecules, and remaining carbon is all locked into dust grains), we can estimate the optical depth of the shell in optical wavelengths, due to the newly-formed carbon grains \citep[see][and references therein]{maeda2013}; 
\begin{eqnarray}
  \tau_{\rm d} & \sim &
  100 \left(\frac{\kappa_{\rm d}}{20,000 \ {\rm cm}^2 \  {\rm g}^{-1}}\right) 
  \left(\frac{X_{\rm d}}{0.1}\right) \left(\frac{M_{\rm env}}{0.1 M_\odot}\right) \nonumber\\
 & &   \left(\frac{V_{\rm shell}}{10,000 \ {\rm km s}^{-1}}\right)^{-2} 
 \left(\frac{t}{200 \ {\rm days}}\right)^{-2} \ ,
\end{eqnarray}
where the dust opacity, $\kappa_{\rm d}$, is scaled to a typical value for carbon grains in optical wavelengths. 

The extinction by the newly-formed dust grains in the late phase is thus generally expected, with some variation depending on the mass of the envelope and possibly the combined effect of the asymmetry and the viewing direction. For example, in case of SN 2020hvf, the relatively small envelope mass (\S 4.1 and Fig. \ref{fig:spec_expand}) could marginally produce a detectable extinction effect for the light through the shocked-envelope region. However, this effect is likely diluted by the non-extinct light through the other region (which has the equal contribution in the example shown in Fig. \ref{fig:spec_expand}). Indeed, in the late-phase, the emission-forming region becomes deeper, and thus the `covering fraction' by the shocked-envelope will decrease for this case, i.e., this behavoir is expected when the line-of-sight connecting the observer and the centre of the explosion does not pass through the envelope. Therefore, as a combination of these effects, we do not expect much extinction in the late-time light curve and spectra of SN 2020hvf. 

As the opposite extreme, a system with the envelope mass of $0.1 M_\odot$ as viewed from the envelope direction could provide a reasonable explanation for SNe 2009dc and 2020esm. In this case, the optical depth by the dust grains will be sufficient to produce a high extinction for these objects. Further, in this geometrical configuration, the shrinking of the emitting region in the velocity space will indeed increase the covering fraction as the emitting region is likely hidden by the overlying envelope along the line-of-sight. The accelerated fading and the suppression of the fluxes in the shorter wavelengths were reported for SNe 2009dc \citep{taubenberger2013} and 2020esm \citep{dimitriadis2022} in the late phase, for which the present scenario provides a possible interpretation. We note that SN 2012dn also shares similar late-time behaviour \citep{taubenberger2019}; the properties of the envelope in our scenario for SN 2012dn are not very different from those for SNe 2009dc and 2020esm, and thus this can be consistent with the present scenario.

\subsection{Two (or more) parameters to characterize the nature of over-luminous SNe Ia?}

As a demonstration, we have shown that the sequence of SN 2020hvf, 2012dn, and then 2020esm/2009dc could be related to an increasing amount of the C+O-rich envelope. This corresponds to the observational sequence of an increasing strength of the C II (see Fig. \ref{fig:spec}), and thus this conclusion might be taken as a straightforward interpretation. It is interesting to note that they also form a sequence of a decreasing velocity in the Si II absorption minimum (Tab. \ref{tab:velocity}). Indeed, this is a tendency not limited in the sample presented here, but seen in a larger sample \citep{ashall2021}. It is also related to the light curve evolution, where the slower decliners are related to stronger C II and slower Si II lines \citep{ashall2021}.

\begin{table}
\centering
\caption{Si II Velocity difference with respect to that in SN 2020hvf:$^{a}$The values for the maximum-light spectra. $^{b}$The values required to shift the model computed for SN 2020hvf. 
}
\label{tab:velocity}
\begin{tabular}{ccc} 
\hline
SN & Observation$^{a}$ & Model$^{b}$ \\
& km s$^{-1}$ & km s$^{-1}$\\
\hline
2012dn & -2,000 & -3,000\\
2009dc & -4,500 & -3,900\\
2020esm & -4.000 & -1,500\\
\hline
\end{tabular}
\end{table}

Given the increasing importance of the deceleration of the ejecta by a larger amount of the envelope (Fig. \ref{fig:density}), the envelope-interaction scenario provides an interesting possibility that the amount of the envelope might provide a rule to explain some of the observational correlations \citep[see also][]{ashall2021}. This, however, is able to explain these correlations at most partly (Tab. \ref{tab:velocity}). If this deceleration effect alone would explain the observed trend in the velocity, one might expect that the line velocities in the model spectra would match to different SNe by simply changing $M_{\rm env}$. We however need to introduce additional redshift for different SNe, starting with the model for SN 2020hvf, to roughly match to the Si II absorption minima of the other SNe. The degree of the additional shift required is different for different SNe; for SN 2020esm, indeed a large part of the observed shift could be attributed to the deceleration effect given the relatively massive envelope considered ($0.1 M_\odot$); on the other hand, for SNe 2012dn and 2009dc, the deceleration by the envelope can account for only a minor fraction of the observed shift. 

Indeed, the envelope, in the parameter space examined here (up to $0.3 M_\odot$), would not much affect the maximum-light photometric properties (Section 3). The peak luminosity has a diversity among the over-luminous SNe Ia, and thus this alone indicates that an additional factor is required to explain the diversity of the over-luminous SNe Ia. Most naturally, the nature of the ejecta may be a controlling factor for the peak properties; this may for example be determined by the mass of the progenitor WD within the thermonuclear explosion scenario. We note that there are weak correlations between the strength of the C II (related to the envelope in the present scenario) and the peak light-curve properties \citep{ashall2021}, which however does not necessarily require that the light-curve diversity is directly caused by the variation in the envelope properties; it might indeed indicate that there could be a relation between the intrinsic ejecta properties (e.g., the WD mass) and the envelope properties (e.g., the envelope mass) that indirectly recovers the observed weak correlation between the C II strength and the peak properties. This might then have important implications for the progenitor scenario(s). 

Another factor that might affect the maximum-light properties is a combination of the asymmetry and viewing angle effects (Section 4.1). If the disc-like envelope structure is considered, the effect of the deceleration will be more substantial to an edge-on observer than to a pole-on observer. This effect may, for example, contribute to some of the differences between SNe 2009dc and 2020esm. 

\subsection{A possible evolutionary channel}
The present work suggests that the C+O-rich envelope with the mass up to $\sim 0.1 M_\odot$ can explain some of the characteristic properties of the over-luminous SNe Ia, i.e., the C II formation and the late-time behaviour (and the initial flash observed for SNe 2020hvf, 2021zny, and 2022ilv, as well as for LSQ12gpw and ASASSN-15pz; see Section 1), in a unified manner. The envelope alone does not affect much of the peak photometric properties (e.g., the slow evolution and the high luminosity), for which an explosion of a super-Chandrasekhar-mass WD has been intensively discussed as one possibility. Evolutionary scenarios to form a super-Chandrasekhar-mass WD have been considered by various researchers. The most popular idea is to introduce a rapidly (and differentially) spinning WD so that the excessive mass can be supported by the centrifugal force. This may be realized both in the SD scenario \citep{liu2010,meng2011,hachisu2012} and in the merging WDs involved in the DD scenario \citep{dan2014,moll2014}. 

The envelope may be associated with either the mass ejection from or accretion on to the progenitor WD. The C+O-rich composition favors a scenario involving a secondary WD (i.e., DD) rather than a non-degenerate companion (i.e., SD). Within the DD scenario, a possible origin of the envelope attached to a super-Chandrasekhar-mass WD might be associated with the disrupted companion WD, as speculated below.  

Let us consider a merger of binary WDs whose total mass exceeds the Chandrasekhar limit. Immediately after the merger, a primary WD will be surrounded by a rapidly rotating torus (or more likely a hot envelope, noting that this `envelope' may not survive until the time of the explosion; see below) as a relic of the tidally-disrupted companion WD. If the angular momentum loss might be sufficiently slow, the accretion of the torus/envelope will keep the primary WD at the critical rotation state \citep{paczynski1991}, which accretes the envelope material beyond the classical Chandrasekhar limit \citep{uenishi2003,yoon2005}. The hot envelope may eventually be lost either due to the accretion onto the primary or mass-loss from the system. 

A possible outcome of this scenario is a rapidly-rotating, massive WD (say, $\sim 2 M_\odot$) supported partly by the centrifugal force. The system then will experience long-term, secular evolution. As the angular momentum is lost from the system, the system may try to keep the hydrodynamic balance by readjusting the angular momentum distribution which may likely be followed by ejection of some amount of the outer material to the equatorial plane. The ejected material is accumulated as an envelope which is to survive until the time of the explosion. At some point, once the angular momentum redistribution does not catch up with the angular momentum loss, then the primary WD will no more support its mass, which will then collapse and explode by thermonuclear runaway initiated by the carbon reactions around the centre. At this point, it is surrounded by the envelope probably confined along the equatorial plane. 

The scenario presented here is highly speculative for the moment, and some variants may also be possible. For example, the geometrical configuration in the resulting envelope in this scenario may also be largely spherical, depending on the angular momentum-loss process; it may also be possible that the `first' hot envelope as a direct relic of the disrupted companion WD (which may be more like spherically distributed after the viscosity-driven redistribution) may indeed survive at the time of the explosion and contribute to the envelope material, rather than considering the `second' envelope ejected during the angular momentum-loss phase. We plan to investigate the scenario(s) further with the stellar evolution calculations in the future. 

\subsection{Implications for other subclasses?}
In the present work, we focus on the application of the scenario to a particular class of over-luminous SNe Ia, specifically adopting massive ejecta interacting with the C+O-rich envelope. However, a similar configuration, i.e., an exploding WD surrounded by an envelope, may be realized in a variety of progenitor/explosion channels. Thus, the general properties predicted in the present work can in principle be applicable to test such scenarios, through the existence of an envelope, for different subclasses of SNe Ia. 

General properties we expect for the C+O-rich envelope, irrespective of details of the nature of the ejecta, can be summarized as follows; 
\begin{enumerate}
    \item Initial flash within the first few days, which is especially strong in the UV wavelengths \citep[see also][]{piro2016,maeda2018}.
    \item For $M_{\rm env} \gsim 0.03 M_\odot$, strong C II lines especially in the earlier phase. 
    \item For $M_{\rm env} \gsim 0.03 M_\odot$, potentially accelerated fading in the late phase. 
\end{enumerate}

The initial flash has been reported for an increasing number of SNe Ia covering different subclasses, but this itself is expected by a number of different mechanisms and interpretation can be difficult \citep[see][for a review]{maeda2018}. On the other hand, the other properties are unique for the C+O-rich envelope. The detection of the C II lines are quite common in normal SNe Ia in the earliest phase \citep{thomas2011,folatelli2012}. The strength of the C II lines are, however, not as strong as those seen in the over-luminous events. Also, normal SNe Ia show no sign of the accelerated luminosity decline in the late phase. These properties could still be accommodated by a small amount of the C+O-rich envelope for normal SNe Ia, up to $\sim 0.03 M_\odot$, while it is either not an evidence of the existence of the envelope. Indeed, by combining the expected properties as summarized above, we can comprehensively test this scenario for SNe Ia with intensive observational coverage from the infant to late phases; the present work thus provides one strong motivation for the high-cadence survey and prompt follow-up observation for SNe Ia of various subclasses, especially nearby events that allow long-term monitoring toward the late phase. In addition, as suggested in Section 4.1 for over-luminous SNe Ia, polarization observations in the very early phase may serve as a smoking gun. 

\section{Conclusions}\label{sec:summary}
In the present work, we have studied the effects of the C+O-rich envelope attached to an exploding WD. We have especially focused on the application of the scenario to a class of over-luminous SNe Ia, or `super-Chandrasekhar-mass' SN Ia candidates. The model has particularly been constructed with the constraints and insights obtained for SN 2020hvf. 

We have found that some of the characteristic features of over-luminous SNe Ia can be explained by this scenario; the envelope leads to the initial flash, strong C II lines and their decreasing strength. A rough match to the trend of lower line velocities for those showing stronger C II lines is also found, which links the over-luminous SNe Ia with stronger C II to the system with a more massive envelope; a possible sequence is SN 2020hvf -- SN 2012dn -- SNe 2009dc/2020esm with the increasing envelope mass. The C+O-rich envelope may further provide a site for the dust formation in the late phase ($\gsim 100$ days), and this may provide a possible interpretation of the accelerated decline in the late-time light curves and the suppression of the fluxes in the shorter wavelengths;  degrees of these effects again form the same sequence as a function of the envelope mass.

The coexistence of the C II and Si II lines persists only in a brief time window in our models especially with a small envelope mass; this may be remedied if the envelope is distributed in an asymmetric way (e.g., disc-like), so that light can reach to an observer both directly and through the shocked envelope. This provides an interesting test for the present scenario; polarization, especially in the very early phase. The asymmetric configuration introduces an additional source of the diversity for observational properties of over-luminous SNe Ia, depending on the degree of the asymmetry and the viewing direction. 

However, the variation in the properties of the envelope and the asymmetric configuration would not explain all the diversity seen in over-luminous SNe Ia, e.g., the variation in the peak luminosity. Therefore, an additional function should exist that controls some of the observational properties of these SNe Ia. The mass of the WD is a straightforward possibility, which may indeed have some variation in the context of the super-Chandrasekhar-mass WD explosion scenario. We have discussed a possible evolutionary channel toward the formation and explosion of the super-Chandrasekhar-mass WD; the formation by the accretion of a tidally disrupted companion WD in a merging binary WD system at a critical rotation, then ejection of the surface layer due to the spin-down evolution after the accretion phase. 

An increasing sample, as well as further modeling effort (including both the radiation transfer and stellar evolution), will be required to further test and constrain the scenario presented here, or in general to understand the nature and the origin of over-luminous SNe Ia. The present work provides various diagnostics that can be tested through future observations, especially in the very infant phase and late phase. 

\section*{Acknowledgements}
K.M. acknowledges support from the Japan Society for the Promotion of Science (JSPS) KAKENHI grant JP18H05223, JP20H00174, and JP20H04737. J.J. acknowledges support from JSPS KAKENHI grants JP18J12714, JP19K23456, and JP22K14069. Numerical computations were carried out on Cray XC50 at Center for Computational Astrophysics, National Astronomical Observatory of Japan.

\section*{Data Availability}
The simulated light curves and spectra are available upon request to K.M.



\bibliographystyle{mnras}
\bibliography{sniacsm} 




%
%


\bsp	
\label{lastpage}
\end{document}